\documentclass[prl,
showpacs,
twocolumn,
dvips,
aps]{revtex4}
\usepackage{amsmath,amssymb,amsfonts,graphicx}

\begin{document}

\title{Heat exchange between two interacting nanoparticles beyond the fluctuation-dissipation regime}
\author{Agustin P\'{e}rez-Madrid$^1$}
\email{agustiperezmadrid@ub.edu}
\author{Luciano C. Lapas$^2$}
\email{lapas@unb.br}
\author{J. Miguel Rub\'{\i}$^1$}
\email{mrubi@ub.edu}
\affiliation{$^1$ Departament de F\'{\i}sica Fonamental, Facultat de F\'{\i}sica, Universitat de Barcelona, Av. Diagonal 647, 08028 Barcelona, Spain.\\
             $^2$ Instituto de F\'{\i}sica and Centro Internacional de F\'{\i}sica da Mat\'{e}ria Condensada, Universidade de Bras\'{\i}lia, Caixa Postal 04513, 70919-970 Bras\'{\i}lia, Distrito Federal, Brazil}

\begin{abstract}
We show that the observed non-monotonic behavior of the thermal conductance between two nanoparticles when they are brought into contact is originated by an intricate phase space dynamics. Here it is assumed that this dynamics results from the thermally activated jumping through a rough energy landscape. A hierarchy of relaxation times plays the key role in the description of this complex phase space behaviour. Our theory enables us to analyze the heat transfer just before and at the moment of contact.
\end{abstract}

\maketitle

\emph{1. Introduction.}--- The knowledge as to how heat transfer takes place at interfaces between different materials in nanoscale systems is an issue which has received considerable attention recently. This phenomenon, involved in high-precision lab techniques as well as in the field of active nanotechnology research~\cite{Domingues05,Various1}, is poorly understood and the theoretical bases behind it are as of yet sketchy. Despite the difficulty to demonstrate the near-field enhancement of radiative heat transfer, many experiments have been performed in this area such as the archetypical of the heat transfer between a nano-tip and a surface~\cite{Various2}. As a consequence of the near-field radiative interactions, transitions among the energy states of the nanoparticles (NPs) related with structural rearrangements occur. This phenomenon is similar to the F\"{o}rster energy transfer~\cite{Forster48}.

The current literature on the subject of radiative energy exchange at the nanoscale is implicitly or explicitly based on the validity of the fluctuation-dissipation theorem~\cite{Joulain05} involving fluctuating electric dipoles. In the particular case of two NPs which are brought together, the near-field radiative energy transfer between them depends greatly on their characteristic size $D$ and separation distance $d$. When the space between them is small, {\em i.e.} $d$ around $2D$, the energy exchange can be considered as due to multipolar interactions in the fluctuation-dissipation regime~\cite{Madrid08} related to exponential or Debye relaxation. Nonetheless, when the NPs get even closer to each other up to contact~\cite{Domingues05}, $d$ near to $D$, the interaction energy becomes of the same order of magnitude as the unperturbed energy of the NPs which lose any crystalline structure they may have. This suggests that the NPs are trapped in a rough energy landscape subjected to a kind of glassy behavior common to a great variety of complex systems such as glasses~\cite{Rubi97} and proteins~\cite{Frauenfelder91,Onuchic97}, just to mention some examples.  In such situations a general methodology is lacking. Characterizing the energy landscape in terms of an order parameter, the heat exchange process is related to the diffusion of non-interacting quasi-particles through the order parameter space. Besides, systems displaying glassy behavior have a multitude of metastable states that they visit after overcoming an irregular distribution of energy barriers. These systems exhibit aging and memory effects inherent to an activated dynamics characterized by a hierarchy of relaxation times~\cite{Madrid03}. Therefore, the NPs cannot be treated as a thermodynamic system at equilibrium and consequently, the fluctuation-dissipation theorem cannot be invoked to explain the energy transfer nor is a multipolar expansion applicable. Then, the question arises as to how to understand the heat exchange between two bodies separated by a nanometric distance in such extreme and nevertheless common cases in experiments.

In this Letter, we have provided a general theory to describe the radiative heat exchange in a general situation. The observed non-monotonic behavior of the thermal conductance between two NPs when they are brought into contact, does suggests an activated process frequently found in a wide range of situations corresponding to complex systems~\cite{Various4}. The nontrivial topology of their phase space is inherent to the rough energy landscape arising from the internanoparticle interaction. In the framework of mesoscopic nonequilibrium thermodynamics~\cite{Vilar01} we find an expression of the radiative heat flux and derive the heat conductance which depends on the Fourier transform of a hierarchy of relaxation times. We then compare our result with previous molecular dynamic data~\cite{Domingues05} reproducing the strong enhancement and subsequent decay of the thermal conductance observed when the NPs are in contact. In our approach the heat exchange is interpreted as a current of quasi-particles seen as stationary waves bounded by both NPs which gives a result in good agreement with the simulation.

\emph{2. Mesoscopic nonequilibrium thermodynamic analysis.}--- We will follow here the classical approach of the activated dynamics in which the system diffuses by thermal hopping through a rough energy landscape. The landscape structures are determined by the strength of the interactions between the microscopic constituents and with external agents, both global and local.

Heat exchange by near-field radiation is based on Coulomb interaction. As a consequence of this interaction,  transitions occur among the energy states of the NPs related with structural rearrangements. Hence, let us assume that the energy exchange process is given by a diffusion current of quanta of energy, quasi-particles~\cite{Landau80b} emitted and absorbed by bodies at different temperatures. The irreversible process represented by the existence of this current can be analyzed in the framework of mesoscopic nonequilibrium thermodynamics based on the assumption of the validity of the second law in the phase space. Hence, let us consider a nonequilibrium gas of quanta characterized by the probability density $\rho (\mathbf{\Gamma,}t)$, where $\mathbf{\Gamma }=(p,x)$ and $p$, $x$ are the momentum and position of a quasi-particle, respectively. According to the Gibbs entropy postulate~\cite{deGroot84,Kampen90}, the density functional
\begin{equation}
S(t)=-k_{B}\int \rho (\mathbf{\Gamma },t)\ln \frac{\rho (\mathbf{\Gamma },t)}{\rho _{eq.}(\mathbf{\Gamma })}d\mathbf{\Gamma +}S_{eq.} \label{entropy}
\end{equation}
is the nonequilibrium entropy of the system, where $S_{eq.}$ is the equilibrium entropy of the gas plus the thermal bath and $\rho_{eq.}(\mathbf{\Gamma })$ is the equilibrium probability density function. Changes in the entropy are related to changes in the probability density which, since the probability is conserved, are given through the continuity equation
\begin{equation}
\frac{\partial }{\partial t}\rho (\mathbf{\Gamma },t)=-\frac{\partial }{\partial \mathbf{\Gamma }}\cdot \mathbf{J}\left( \mathbf{\Gamma },t\right) \label{continuity}
\end{equation}
where $\partial /\partial \mathbf{\Gamma }=(\partial /\partial p,\nabla )$, $\nabla =\partial /\partial x$. The continuity equation (\ref{continuity})
defines the probability current $\mathbf{J}\left( \mathbf{\Gamma ,}t\right)=(J_{x},J_{p})$ to be determined after computing the entropy production which follows by combining the time derivative of Eq. (\ref{entropy}) and (\ref{continuity}) to give
\begin{equation}
\frac{\partial S}{\partial t}=-\int \mathbf{J}\left( \mathbf{\Gamma ,}t\right) \cdot \frac{\partial }{\partial \mathbf{\Gamma }}\frac{\mu (\mathbf{\Gamma ,}t)}{T}d\mathbf{\Gamma }\geq 0\text{.}\label{entropy-prod}
\end{equation}
This is an equation which according to the second law and analogous to the form of the entropy production in macroscopic diffusion~\cite{deGroot84}, expresses the entropy production in terms of the nonnegative product of a thermodynamic current $\mathbf{J}\left( \mathbf{\Gamma ,}t\right)$ and its conjugated thermodynamic force $\partial /\partial \mathbf{\Gamma }\left[\mu (\mathbf{\Gamma ,}t)/T\right] $, the gradient of the nonequilibrium chemical potential
\begin{equation}
\mu (\mathbf{\Gamma },t)=k_{B}T\ln \frac{\rho (\mathbf{\Gamma },t)}{\rho_{eq.}(\mathbf{\Gamma })} \label{chem-pot}
\end{equation}
which derives directly from Eq. (\ref{entropy}). The physical meaning of the nonequilibrium chemical potential (\ref{chem-pot}) becomes clearer when we rewrite Eq. (\ref{entropy})
\begin{equation}
\delta S=-\frac{1}{T}\int \mu (\mathbf{\Gamma ,}t)\delta\rho(\mathbf{\Gamma ,}t)d\mathbf{\Gamma }
\end{equation}
which is the Gibb's equation in the phase space. Since $-T\delta S=\delta A$, with $A$ the nonequilibrium free energy, the nonequilibrium chemical potential (\ref{chem-pot}) can be interpreted as the free energy per unit of probability mass. These concepts are also common in other branches of condensed-matter physics ~\cite{gennes}.

The entropy production constitutes the nonequilibrium potential of the system which not far from equilibrium allows us to infer regression laws relating currents and the conjugated thermodynamic forces. Thus in our case, from Eq. (\ref{entropy-prod}) we find
\begin{equation}
\mathbf{J}\left( \mathbf{\Gamma ,}t\right) =-\mathbf{L}\cdot \frac{\partial}{\partial \mathbf{\Gamma }}\frac{\mu (\mathbf{\Gamma ,}t)}{T}\text{,}
\label{pheno1}
\end{equation}
with $\mathbf{L}\left( \rho \right)$ being the matrix of Onsager coefficients which, as required for the second law, should be positive-definite. By combining here Eqs. (\ref{chem-pot}) and (\ref{pheno1}) we obtain
\begin{equation}
\mathbf{J}\left( \mathbf{\Gamma },t\right) =-\mathbf{D}(\rho )\cdot \frac{\partial }{\partial \mathbf{\Gamma }}\rho \text{,} \label{pheno2}
\end{equation}
where $\mathbf{D}(\rho )=k_{B}\mathbf{L}/\rho $ is the matrix of diffusion coefficients. Whence, Eq. (\ref{pheno2}) yields
\begin{equation}
\left\{
\begin{array}{c}
J_{p}=-D_{pp}\dfrac{\partial }{\partial p}\rho -D_{px}\dfrac{\partial }{\partial x}\rho \\
\\
J_{x}=-D_{xp}\dfrac{\partial }{\partial p}\rho -D_{xx}\dfrac{\partial }{\partial x}\rho
\end{array}
\right. \text{.} \label{pheno3}
\end{equation}

Since the quanta are massless particles, there is no diffusion in physical or real space, $J_{x}=0$. Therefore, from the set of equations (\ref{pheno3}) we obtain the diffusion current in momentum space
\begin{equation}
J_{p}=\left( \frac{D_{pp}D_{xx}}{D_{xp}}-D_{px}\right) \frac{\partial }{\partial x}\rho \equiv \frac{\hbar }{\tau (\rho )}\frac{\partial }{\partial x}\rho  \label{momen-curr}
\end{equation}
which defines  $\tau (\rho )$, the relaxation time  simply related to the effective diffusion along the reaction coordinate $x$. Additionally,  the size of an elementary cell in the phase space is given by $\hbar$ and, since according to the theory of Brownian motion the diffusion coefficient $D$, basically the mean square displacement (an area) per unit of time, is in direct relationship to the surface visited in the unit time in the phase space, we can therefore interpret $D$ as the ratio between the size of a characteristic cell,  $\hbar$, and a characteristic time, $\tau (\rho )$.

By integrating Eq. (\ref{momen-curr}) through from $x_{1}$ to $x_{2}$, the position of both NPs, respectively, we find the net current
\begin{equation}
J=\frac{\hbar }{\tau ^{\ast }}\left( \rho _{2}-\rho _{1}\right) \text{,}\label{relax}
\end{equation}
where $\rho _{j}(p,t)=$ $\rho (p,x=j,t)$ is understood as related to the population of quasi-particles at $x_{j}$. Here, we have defined the net
current as
\begin{equation}
J \left(p,t\right) \equiv \frac{1}{\tau^{\ast }\left( t\right)} \int_{1}^{2} \tau (\rho ) J_{p}dx \label{tauast}
\end{equation}
where 
\begin{equation}
\tau ^{\ast }\left( t\right) =\int \rho \;\tau \left( \rho \right) d\mathbf{\Gamma } \label{tauast2}
\end{equation}
stands for a hierarchy of relaxation times corresponding to a non-Debye process proper of the dynamics of complex systems~\cite{Madrid05}.

\emph{3. Stationary state and thermal conductivity.}--- At equilibrium $\rho_{2}=\rho _{1}$ and $J=0$. However, if we maintain the system in a stationary state in which each one of the NPs remains in local equilibrium with its respective bath at different temperatures so that $\rho _{j}\rightarrow \rho _{j,eq.}(T_{j})$ with 
\begin{equation}
\rho _{j,eq.}(T_{j})=\frac{\exp \left( -\beta _{j}E_{n}\right) }{\sum_{n}\exp \left( -\beta _{j}E_{n}\right) } \label{canonical}
\end{equation}
being the canonical distribution, where $\beta _{j}=1/k_{B}T_{j}$ and $E_{n}$ refers to the energy levels of the NP, then the current given through Eq. (\ref{relax}) reaches a non-zero stationary value,
\begin{equation}
J_{st}=\frac{\hbar }{\tau ^{\ast }}\left[ \rho _{1,eq.}(T_{1})-\rho_{2,eq.}(T_{2})\right] \text{.} \label{current-2}
\end{equation}
The fact that each nanoparticle remains in local equilibrium with its proper bath corresponds with one among several possible metastable states of the composed system. 

We find the energy flux by multiplying Eq. (\ref{current-2}) through by $E_{n}$ and canonically averaging, yielding
\begin{equation}
Q(\omega)=\frac{\hbar}{\tau^{\ast}(\omega)}\left[ \Theta(\omega,T_{1})-\Theta(\omega,T_{2})\right] \text{,} \label{energy-flow}
\end{equation}
where we have defined the frequency $\omega=\Delta/\hbar$, with $\Delta$ representing the gap between energy levels, $\Theta(\omega,T)=\hslash\omega N(\omega,T)$ is the average energy of an harmonic oscillator and $N(\omega,T)=1/\left( \exp (\hslash\omega/k_{B}T ) - 1\right) $ is the Planck distribution. The heat conductance follows by linearizing Eq. (\ref{energy-flow}) with respect to the temperature difference:
\begin{equation}
G(\omega,T_{o})=\frac{k_{B}\hbar}{4\tau^{\ast}(\omega)}\left( \frac {\hbar\omega/k_{B}T_{o}}{\sinh(\hslash\omega/2k_{B}T_{o})}\right)^{2} \text{,} \label{conductance-2}
\end{equation}
where $T_{o}=(T_{1}+T_{2})/2$ is the temperature corresponding to the stationary state of the system. It should be noted that these two spheres of the same diameter maintained at different temperatures constitute a model corresponding to more realistic systems such as for example a hot tip in contact with a flat substrate, as mentioned above. When $d=2\pi c/\omega$, a phonon-like dispersion relation, Eq. (\ref{conductance-2}) gives us the heat conductance as a function of the distance between the NPs. Here, the effective relaxation time $\tau^{\ast}(\omega)$ plays the role of an adjustable parameter which in general depends on the frequency according to the fact that for extremely close distances the system adopts a glassy behavior typical of complex systems.
\begin{figure}[!ht]
\includegraphics[scale=0.65]{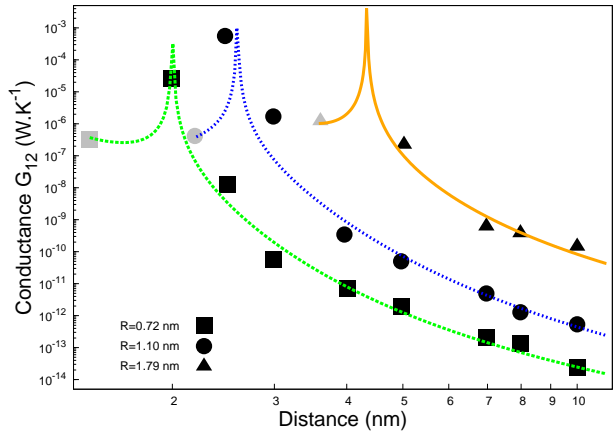}
\caption{(Color online) Thermal conductance $G_{12}$ vs distance $d$
reproducing the molecular dynamics data obtained by Domingues \textit{et al.}~\cite{Domingues05}. The grey points represent the conductance when the particles with effective radius $R=0.72$, $1.10$, and $1.79$ nanometers are in contact. The lines show the analytical result obtained from Eq. (\ref{conductance-2}) by adjusting $\tau^{\ast}(\omega)$ to the simulation data.}
\label{fig2}
\end{figure}

The intricate slow dynamics which complex systems exhibit is represented by a hierarchy of relaxation times  often given by a Kohlrausch-Williams-Watts law~\cite{Various3}

\begin{equation}
\tau ^{\ast }(t)=a\exp\left( -bt^{\beta }\right) \text{, with}\; 0<\beta\leq 2\text{,} \label{tau}
\end{equation}
also known as stretched exponential and where $\beta$ is a matching parameter and $a$ and $b$ are constants which depend on the size of the NPs. This is an empirical time-law, ubiquitous in disordered systems such as glasses which can be related to the distribution of residence times or the response function \cite{Madrid05,Vainstein06}. Therefore, in our case we have tried a time law given by a stretched exponential to make Eq. (\ref{conductance-2}) fit the available numerical simulation data~\cite{Domingues05}. We have found that an exponent $\beta =2$ for which $\tau^{\ast }(t)$ as well as its Fourier transform $\tau ^{\ast }(\omega )$ are Gaussians, gives an excellent fit. This is supported by Fig. (\ref{fig2}), where we have represented the heat conductance as a function of the distance $d$ between the NPs of different effective radii $R=D/2$. This figure shows a strong enhancement of the heat conductance when $d$ decreases until around $2D$ nm due to multipolar interactions~\cite{Madrid08}. When both NPs are in contact, a sharp fall occurs which can be interpreted as due to an intricate conglomerate of energy barriers inherent to the amorphous character of these NPs generated by the strong interaction. The behavior shown in Fig. (\ref{fig2}) agrees with what Eq. (\ref{conductance-2}) predicts since in this equation, the two factors $k_{B}\hbar /4\tau ^{\ast }(\omega )$ and $\{ (\hbar \omega k_{B}T_{o})/\sinh (\hbar \omega /2k_{B}T_{o})\} ^{2}$ compete. For large distances the dominant factor is the latter factor whereas at contact the former factor becomes the leading term. For large length scales, when $\tau ^{\ast }$ is almost a constant which corresponds with exponential relaxation the fluctuation-dissipation theorem is satisfied and we find the system in the fluctuation-dissipation regime. Therefore, the conclusions of our theory are clearly well supported by Fig. (\ref{fig2}) which indicates that the hypothesis of the rough energy landscape is indeed reasonable.

As far as we know the decay in the thermal conductivity at contact is an open question not yet resolved. Here, based on the fact that the crystalline structure of the NPs rapidly disappears when the internanoparticle distance is decreased~\cite{Domingues05}, the assumption that the resulting amorphous material self-organizes into an irregular distribution of energy barriers is the key to interpreting the origin of this decay. Namely, the classical maximum thermal conductance applies to the fluctuation-dissipation regime before contact in which the system preserves its crystalline structure and optical phonons are mainly excited. However, since it has been observed that both NPs become amorphous-like particles when they come closer together due to their strong interaction, we have supposed that they behave as glassy materials. In glasses as well as in other disordered materials, it is a known fact that the distribution of modes presents anomalies which result from an excess of low-frequency modes [20], the so-called boson peak. Hence, it is not possible to make a clear distinction between optical and acoustic modes. These low-frequency modes involve collective motions causing extremely slow structural relaxation. Therefore, this accounts for the high non-linear behavior of the thermal conductance between both NPs.

\emph{4. Conclusion.}--- Up to now nanoscale radiative heat transfer has been studied assuming the validity of the fluctuation-dissipation theorem related to exponential or Debye relaxation. This approach shows itself to be successful to describe the energy exchange between two NPs before contact. At contact, however, since the interaction between both NPs does not admit a multipolar expansion this approach ceases to be valid. Hence, in view of the subsequent amorphous character of both NPs, we have assumed that this interaction gives rise to an irregular array of energy barriers and therefore, the relaxation of both NPs becomes a complex activated process. Thus, the transfer of energy is assumed to be due to the diffusion of quasi-particles through a rough energy landscape which gives rise to activated processes related to a rate current derivable in the framework of our thermodynamic theory. Our theory provides us the expression of the thermal conductance which depends on an adjustable parameter representing a hierarchy of relaxation times,  analogous to the slow relaxation behavior exhibited by similar complex systems. We have found that a time law similar to a stretched exponential with an exponent to the second power perfectly fits the available molecular dynamics simulation data, thus supporting our approach.

This work was supported by the DGiCYT of Spanish Government under Grant No. FIS2008-04386, and by Brazilian Research Foundations: FAPDF, CAPES and CNPq.

\end{document}